\documentclass[runningheads]{llncs}

\usepackage{amsmath}
\usepackage{amsfonts}
\usepackage{amssymb}
\usepackage{graphicx}
\usepackage[colorlinks]{hyperref}
\usepackage[switch]{lineno}
\usepackage[dvipsnames]{xcolor}
\usepackage{tikz}        
\usepackage{forest}      

\graphicspath{{figures/}}

\tikzset{
  solid node/.style={circle,draw,inner sep=1.2,fill=black},
  hollow node/.style={circle,draw,inner sep=1.2},
  left label/.style={above left,midway},
  right label/.style={above right,midway},
  mid label/.style={above,midway}
}

\title{Auditing Ranked Voting Elections with Dirichlet-Tree Models: First
Steps\thanks{To appear at E-Vote-ID 2022.
We thank Ronald Rivest for many helpful suggestions for improving the paper.
This work was supported by the University of Melbourne's Research Computing
Services and the Petascale Campus Initiative; and by the Australian Research
Council (Discovery Project DP220101012).}}

\author{
Floyd Everest     \inst{1}   \orcidID{0000-0002-2726-6736}  \and
Michelle Blom     \inst{2}   \orcidID{0000-0002-0459-9917}  \and
Philip B. Stark   \inst{3}   \orcidID{0000-0002-3771-9604}  \and
Peter J. Stuckey  \inst{4}   \orcidID{0000-0003-2186-0459}  \and
Vanessa Teague    \inst{5,6} \orcidID{0000-0003-2648-2565}  \and
Damjan Vukcevic   \inst{1,7} \orcidID{0000-0001-7780-9586}}

\authorrunning{Everest, Blom, Stark, Stuckey, Teague, Vukcevic}

\institute{
School of Mathematics and Statistics, University of Melbourne, Parkville,
Australia
\and
School of Computing and Information Systems, University of Melbourne,
Parkville, Australia
\and
Department of Statistics, University of California, Berkeley, CA, USA
\and
Department of Data Science and AI, Monash University, Clayton, Australia
\and
Thinking Cybersecurity Pty.\ Ltd., Melbourne, Australia
\and
The Australian National University, Canberra, Australia
\and
Melbourne Integrative Genomics, University of Melbourne, Parkville,
Australia \\
\email{damjan.vukcevic@unimelb.edu.au}}

\begin{document}

\maketitle

\begin{abstract}
Ranked voting systems, such as instant-runoff voting (IRV) and single
transferable vote (STV), are used in many places around the world.  They are
more complex than plurality and scoring rules, presenting a challenge for
auditing their outcomes: there is no known risk-limiting audit (RLA) method for
STV other than a full hand count.

We present a new approach to auditing ranked systems that uses a statistical
model, a Dirichlet-tree, that can cope with high-dimensional parameters in a
computationally efficient manner.  We demonstrate this approach with a
ballot-polling Bayesian audit for IRV elections.  Although the technique is not
known to be risk-limiting, we suggest some strategies that might allow it to be
calibrated to limit risk.
\end{abstract}



\noindent
In \emph{ranked voting}, voters rank candidates in order of preference; some
elections require a complete ranking, others allow partial rankings.  Counting
the votes can be complex, e.g.\ involving potentially long sequences of
eliminations of candidates (for IRV), and transfers of weighted votes between
candidates (for STV).

Complexity arises in two ways: (i)~a very large number of ways to fill out a
ballot ($k!$ ways to rank $k$ candidates); (ii)~the social choice functions are
sensitive, small changes can sometimes drastically alter the outcome.  This
poses a challenge for auditing: we require statistical inference in a very
high-dimensional parameter space, for a function prone to erratic behaviour.

RLAs have been developed for some ranked voting systems: (i)~IRV elections
\cite{blom2019raire}; (ii)~2-seat STV elections \cite{blom2021stv}.  Both RLAs
project into lower dimensions, where statistical testing is tractable.
However, their projections typically capture only a subset of elimination
sequences that lead to the winner.  If the true sequence is not one of those,
but leads to the same winner, then the audits will usually (and unnecessarily)
escalate to a full count despite the reported winner being correct.

Thus, there is scope for further development for ranked systems.  For IRV we
seek a method that can work with a more complete set of elimination sequences,
and for STV we want to be able to audit elections with more than 2
winners.\footnote{%
E.g., Australian Senate elections use STV to elect up to 12 candidates for each
state.}

We tackle the problem directly as a Bayesian audit \cite{rivest2012}.  This is
challenging in high-dimensions; a previous attempt
\cite{chilingirian2016auditing} gave up on fitting a full model and instead
used a bootstrap approach (equivalent to a degenerate Bayesian model).

Our contribution is a new specification of the statistical model that works
efficiently in high-dimensions, making Bayesian audits possible for ranked
voting elections.  We demonstrate this with examples of auditing IRV elections.


\section{Dirichlet-tree model for ranked voting}

An audit involves calculating the evidence in favour of the reported outcome
using a sample of ballots and a statistical model.  For ranked voting, the
natural model is multinomial: each ballot type (ranking of the candidates)
occurs with some (fixed but unknown) frequency across all ballots.

A Bayesian audit can work with this model directly, by specifying a prior
distribution on the ballot probabilities.  Given a sample of ballots, we obtain
a posterior distribution for these probabilities, which induces a posterior
distribution on the winner(s).  If the reported outcome exceeds some desired
posterior probability threshold, we stop the audit, otherwise we sample more
ballots.

For a multinomial model, a typical choice of prior is a Dirichlet distribution.
This is conjugate, allowing convenient and efficient implementation.  It is
defined by concentration parameters, $a_i > 0$, for each ballot type $i \in
\{1, 2, \dots, K\}$.  The posterior is Dirichlet (by conjugacy) with
concentration parameters $a_i + n_i$ after observing $n_i$ ballots of type $i$.
To make the prior candidate-agnostic: $\forall i, a_i = a_0$ for some $a_0$.
Setting $a_0 = 1$ gives a uniform density on the space of probabilities.

This model behaves poorly as $K$ grows very large.  If we set $a_0 = 1$, the
prior becomes very informative: it will swamp the data, making the posterior
converge very slowly.  If we set $a_0$ much smaller, for example $a_0 \approx 1
/ K$, then the posterior will strongly concentrate on the ballot types observed
in the sample, approximating a `bootstrap' method.  This will likely understate
the uncertainty.  It will also be challenging to implement, with values of
$1 / K$ being smaller than typical machine precision once there are about 30
candidates.

To overcome these issues, we propose using a Dirichlet-tree prior distribution
(e.g.\ \cite{dirichlet-tree-distribution}).\footnote{%
Our implementation is available at:
\url{https://github.com/fleverest/elections.dtree}} This is a
set of nested Dirichlet distributions with the nesting described by a tree
structure.  It generalises the Dirichlet while retaining conjugacy with the
multinomial.  The nesting divides up the space, allowing efficient inference in
high dimensions.

\begin{figure}
\centering
\begin{forest}
[ , for tree={s sep=2cm}, hollow node
  [ ,for tree={hollow node},edge label={node[left label]{$a_0$}}
    [ ,for tree={solid node},label={below:$p(1,2,3)$},edge label={node[left label]{$a_0$}}]
    [ ,for tree={solid node},label={below:$p(1,3,2)$},edge label={node[right label]{$a_0$}}]
  ]
  [ ,for tree={hollow node},edge label={node[left label]{$a_0$}}
    [ ,for tree={solid node},label={below:$p(2,1,3)$},edge label={node[left label]{$a_0$}}]
    [ ,for tree={solid node},label={below:$p(2,3,1)$},edge label={node[right label]{$a_0$}}]
  ]
  [ ,for tree={hollow node},edge label={node[left label]{$a_0$}}
    [ ,for tree={solid node},label={below:$p(3,1,2)$},edge label={node[left label]{$a_0$}}]
    [ ,for tree={solid node},label={below:$p(3,2,1)$},edge label={node[right label]{$a_0$}}]
  ]
]
\end{forest}
\caption{Dirichlet-tree prior for ranked voting ballots with 3 candidates.}
\label{fig:irv_tree_prior}
\end{figure}
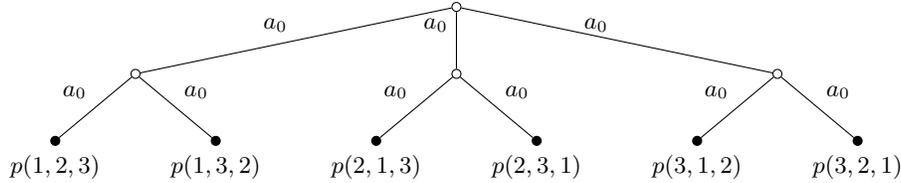

The tree structure we propose follows the preference ordering: the first split
in the tree has a branch for each possible first preference (one branch per
candidate), the next split has a branch for each possible second preference
(amongst remaining candidates), etc.  Partial ballots are modelled by
`termination' branches.  To initialise the prior, we set the concentration
parameter for each branch to be equal to $a_0$; see
\autoref{fig:irv_tree_prior} for an example with no partial ballots.

\section{Ballot-polling Bayesian audits of IRV elections}

We demonstrate our model using data from two elections of different sizes:
(i)~Seat of Albury, NSW 2015 lower house elections, Australia [5 candidates;
46,357 ballots]; (ii)~San Francisco Mayoral election 2007 [18 candidates;
149,465 ballots].\footnote{%
Data source: \url{https://github.com/michelleblom/margin-irv}}
The latter has more than $18! \approx 6.4 \times 10^{15}$ possible ballot
types.

We used a Dirichlet-tree prior that allows partial ballots and had $a_0 = 0, 1,
10, 100$.  We also used a Dirichlet prior with $a_0$ set such that its prior
variance, for an arbitrary complete ballot, matched that of the corresponding
Dirichlet-tree prior.  Setting $a_0 = 0$ for either prior gives a `bootstrap'
audit~\cite{chilingirian2016auditing}.

For each election, we simulated 100 audits by randomly permuting the ballots
(without introducing any errors).  We took samples of up to 200 ballots for
Albury and up to 50 for San Francisco, which was sufficient to illustrate the
differing behaviour of the priors.  At each point in the audit, we estimated
posterior probabilities by taking the mean of 100 draws from the posterior.

\begin{figure}
\centering
\includegraphics[width=\textwidth]{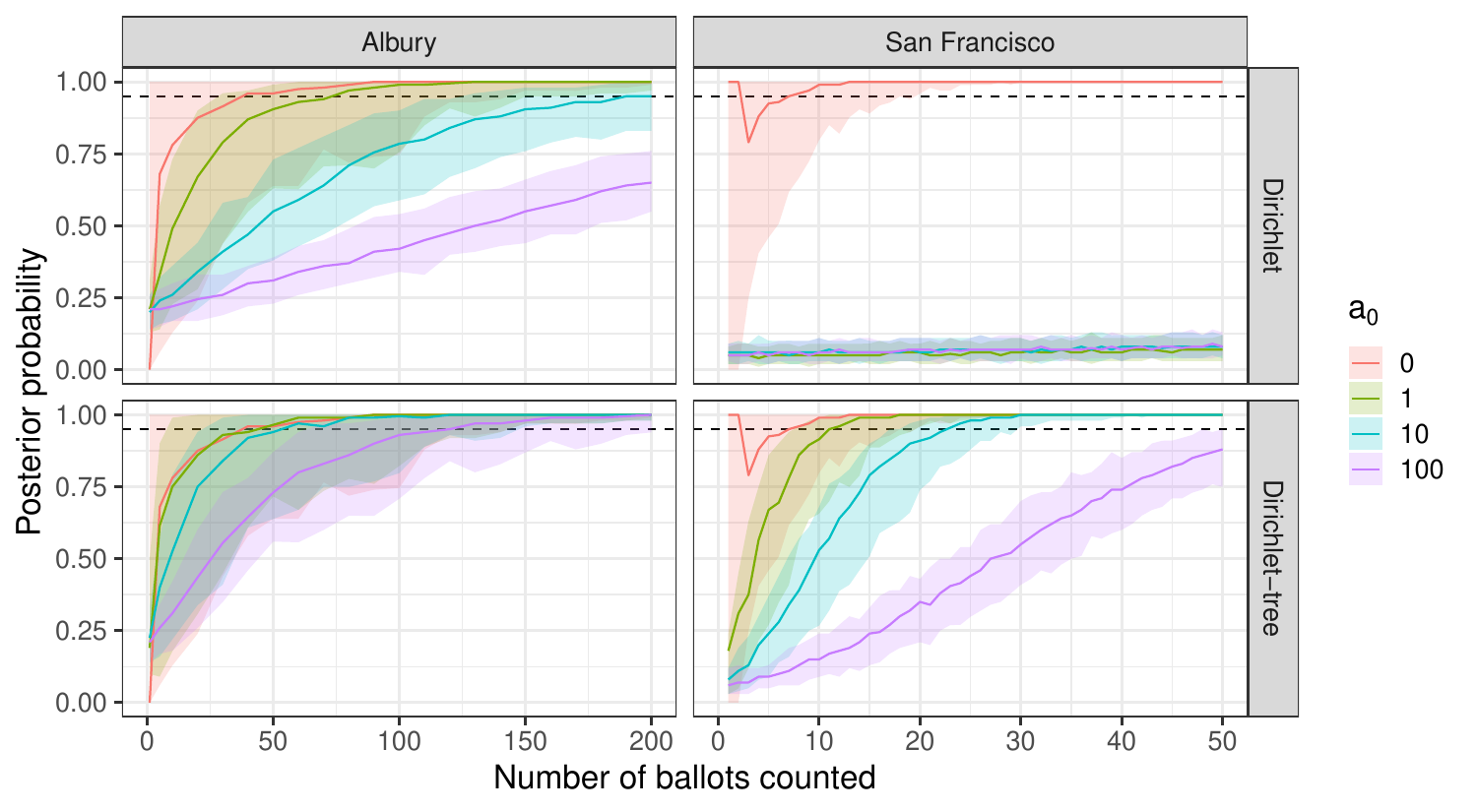}
\caption{Distribution of the posterior probability for the winner, vs sample
size.  The lines show the median across 100 simulated audits, the corresponding
bands shade the values between the 5\% and 95\% quantiles.  The dashed line
shows a posterior probability of 0.95, for reference.  The $a_0$ values refer
to the Dirichlet-tree prior; for the Dirichlet a `corresponding' value was
chosen (see main text).}
\label{fig:posterior-paths}
\end{figure}

\autoref{fig:posterior-paths} shows how the posterior probability for the true
winner evolved as the samples increased.  The Dirichlet-tree model worked for
both elections and responded to $a_0$ as expected: increasing it made the prior
more informative and hence respond more slowly to data.  The Dirichlet model
behaved similarly when we had only a few candidates (Albury) but unstable when
we had many (San Francisco), with all choices except the bootstrap ($a_0 = 0$)
being too informative.

The bootstrap was erratic at the start (a wide range of posterior values) and
stabilised once the sample was big enough.  In practice, the poor
regularisation at the start would lead to increased risk.  Whether this can be
curbed by simply specifying a minimum sample size is worth investigating in
general.

\section{Discussion}

\enlargethispage{\baselineskip}

We have demonstrated a statistical model that allows efficient ballot-polling
Bayesian audits of ranked voting elections.  While our example was specifically
for IRV, the model can be applied to any ranked voting election by simply
changing the social choice function in the calculation of the posterior
distribution.  Furthermore, the tree structure can be adapted to better suit
specific features of particular elections, which should improve efficiency.

A current limitation of our approach is that it cannot be used to run an RLA.
This requires an easy way to compute or impose a risk limit.  We propose two
ways to overcome this: (i)~determine the maximum possible risk by deriving the
worst-case configuration of true ballots, such as was done for 2-candidate
elections \cite{huang_unified_2020}; (ii)~use a prior-posterior ratio (PPR)
martingale \cite{waudby-smith-2020-confseq-wor} to make an RLA using the
Dirichlet-tree model.  Another limitation is that our approach currently only
supports ballot-polling audits.  Adapting it to allow other types of audits,
such as comparison audits, is another important avenue for future work.


\bibliographystyle{splncs04}
\bibliography{references}


\end{document}